\documentclass[preprint,nofootinbib,3p]{elsarticle}
\usepackage{graphicx}
\usepackage{amsmath}
\usepackage{amssymb}
\usepackage{overpic,subfigure}
\usepackage{multirow}
\usepackage[symbol]{footmisc}
\usepackage{dcolumn}
\usepackage{bm}
\usepackage{rotating}
\usepackage{hyperref}
\hypersetup{backref,
pdfpagemode=FullScreen,
colorlinks=true}
\usepackage{indentfirst}
\usepackage{lineno}
\usepackage{epstopdf}
\usepackage{units}
\usepackage{amsthm}
\usepackage{url}
\usepackage{amsfonts}
\usepackage{textcomp}
\usepackage{multicol}
\usepackage{verbatim}
\usepackage{rotating}
\usepackage{float}
\usepackage{color}
\usepackage{pstricks}
\usepackage{pst-node}
\usepackage{times}
\usepackage{ulem}
\usepackage{booktabs}
\usepackage{multicol}

\newcommand{\dedx}{dE/dx}
\newcommand{\BR}{{\cal B}}

\newcommand{\pip}{\pi^+}

\newcommand{\piz}{\pi^0}

\newcommand{\ppm}{\pi^{\pm}}
\newcommand{\kpm}{K^{\pm}}
\newcommand{\bfg}{\begin{figure}}
\newcommand{\efg}{\end{figure}}
\newcommand{\bitm}{\begin{itemize}}
\newcommand{\eitm}{\end{itemize}}
\newcommand{\bnum}{\begin{enumerate}}
\newcommand{\enum}{\end{enumerate}}
\newcommand{\btbl}{\begin{table}}
\newcommand{\etbl}{\end{table}}
\newcommand{\btbu}{\begin{tabular}}
\newcommand{\etbu}{\end{tabular}}

\newcommand{\beq}{\begin{equation}}
\newcommand{\edq}{\end{equation}}

\newcommand{\gev}{GeV}

\makeatletter
\newenvironment{tablehere}
  {\def\@captype{table}}
  {}
\newenvironment{figurehere}
  {\def\@captype{figure}}
  {}
\makeatother
\begin{document}
\begin{frontmatter}
\title{\boldmath Measurement of branching fractions for $D$ meson decaying into $\phi$ meson and a pseudoscalar meson}
\author{
\begin{small}
\begin{center}
M.~Ablikim$^{1}$, M.~N.~Achasov$^{10,d}$, P.~Adlarson$^{59}$, S. ~Ahmed$^{15}$, M.~Albrecht$^{4}$, M.~Alekseev$^{58A,58C}$, A.~Amoroso$^{58A,58C}$, F.~F.~An$^{1}$, Q.~An$^{55,43}$, Y.~Bai$^{42}$
% ...
\\
\vspace{0.2cm}
(BESIII Collaboration)\\
\vspace{0.2cm} {\it
$^{1}$ Institute of High Energy Physics, Beijing 100049, People's Republic of China\\
% ...
$^{4}$ Bochum Ruhr-University, D-44780 Bochum, Germany\\
% ...
$^{10}$ G.I. Budker Institute of Nuclear Physics SB RAS (BINP), Novosibirsk 630090, Russia\\
% ...
$^{15}$ Helmholtz Institute Mainz, Johann-Joachim-Becher-Weg 45, D-55099 Mainz, Germany\\
% ...
$^{42}$ Southeast University, Nanjing 211100, People's Republic of China\\
$^{43}$ State Key Laboratory of Particle Detection and Electronics, Beijing 100049, Hefei 230026, People's Republic of China\\
% ...
$^{55}$ University of Science and Technology of China, Hefei 230026, People's Republic of China\\
% ...
$^{58}$ (A)University of Turin, I-10125, Turin, Italy; (B)University of Eastern Piedmont, I-15121, Alessandria, Italy; (C)INFN, I-10125, Turin, Italy\\
$^{59}$ Uppsala University, Box 516, SE-75120 Uppsala, Sweden\\
% ...
$^{d}$ Also at the Novosibirsk State University, Novosibirsk, 630090, Russia\\
% ...
}\end{center}
\vspace{0.4cm}
\end{small}
}
\date{\today}
\begin{abstract}

 The four decay modes $D^{0}\to\phi\pi^{0}$, $D^{0}\to\phi\eta$, $D^{+}\to\phi\pi^{+}$, and $D^{+}\to\phi K^{+}$ are studied by using a data sample taken at the centre-of-mass energy $\sqrt{s} = 3.773$~GeV~with the BESIII detector, corresponding to an integrated luminosity of 2.93 fb$^{-1}$. The branching fractions of the first three decay modes are measured to be
$\BR(D^{0}\to\phi\pi^{0})=(1.168\pm0.028\pm0.028)\times10^{-3}$,
$\BR(D^{0}\to\phi\eta)=(1.81\pm0.46\pm0.06)\times10^{-4}$, and
$\BR(D^{+}\to\phi\pi^{+})=(5.70\pm0.05\pm0.13)\times10^{-3}$, respectively, where the first uncertainties are statistical and the second are systematic. In addition, the upper limit of the branching fraction for $D^+\to\phi{K^+}$ is given to be $2.1\times10^{-5}$ at the $90\%$ confidence level. The ratio of $\BR(D^0\to\phi\pi^0)$ to $\BR(D^+\to\phi\pi^+)$ is calculated to be $(20.49\pm0.50\pm0.45)\%$, which is consistent with the theoretical prediction based on isospin symmetry between these two decay modes.

\end{abstract}
\end{frontmatter}
\begin{multicols}{2}
\section{Introduction}

\begin{figure*}[tbp]
\centering
\subfigure[Color suppressed $D^+\to\phi\pi^+$]{
\includegraphics[width=4.8cm]{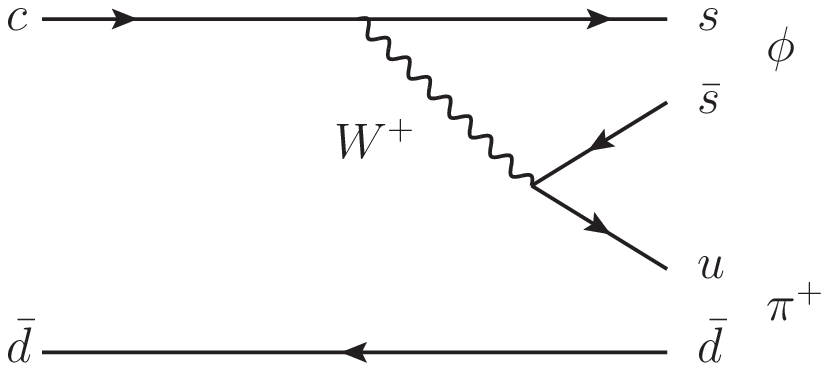}
}
\quad
\subfigure[Color suppressed $D^0\to\phi\pi^{0}$]{
\includegraphics[width=4.8cm]{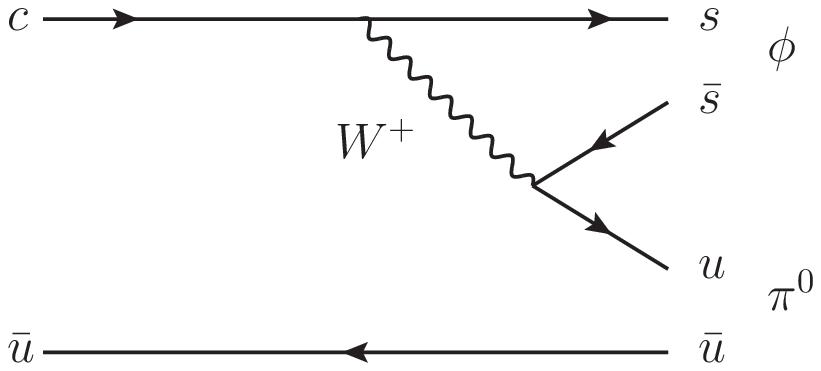}
}
\quad
\subfigure[W-annihilation $D^+\to\phi{K^+}$]{
\includegraphics[width=4.8cm]{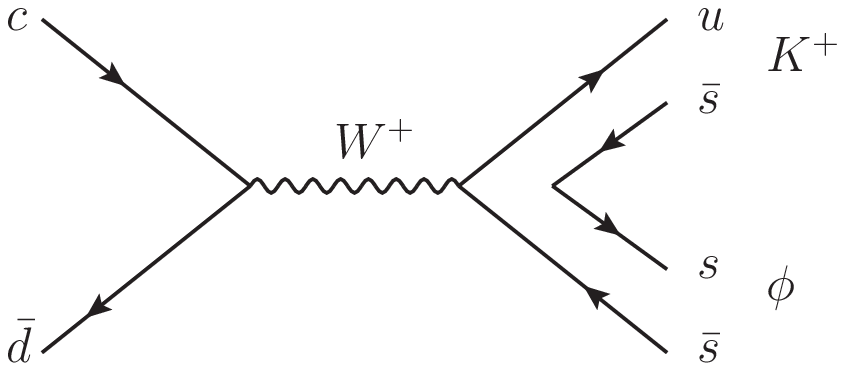}
}
\quad
\subfigure[Color suppressed $D^0\to\phi\eta$]{
\includegraphics[width=4.8cm]{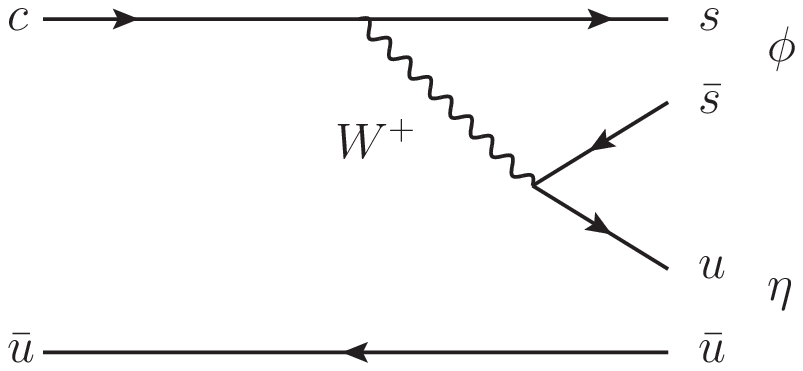}
}
\quad
\subfigure[W-exchange $D^0\to\phi\eta$]{
\includegraphics[width=4.8cm]{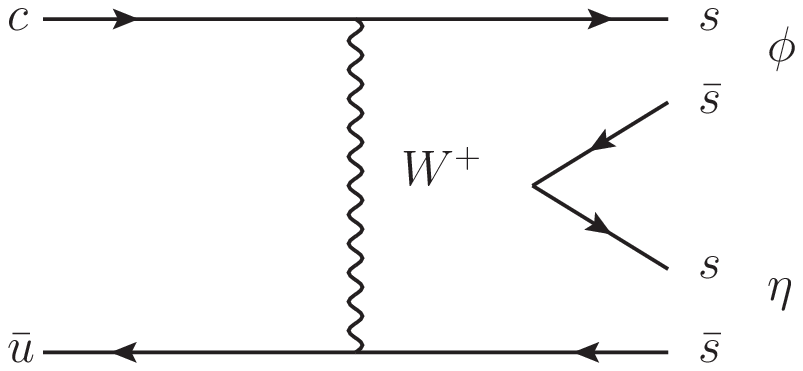}
}
\quad
\subfigure[W-exchange $D^0\to\phi\eta$]{
\includegraphics[width=4.8cm]{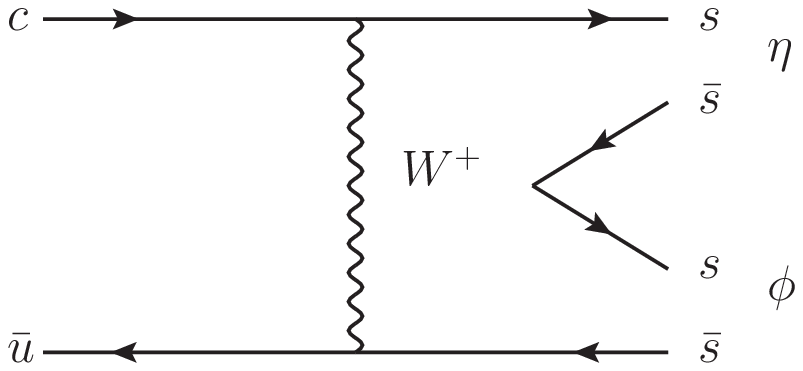}
}
\caption{Feynman diagrams of four $D\to\phi{P}$ decay modes. }
\label{figure:feynman}
\end{figure*}

Comprehensive and precise measurements of hadronic $D$~~meson decays provide important inputs for the experimental studies of both charm and beauty decays~\cite{2009besphys}.
One category of decay modes $D\to\phi{P}$ ($P$ represents a pseudoscalar particle) has simple Feynman diagrams as depicted in Fig.~\ref{figure:feynman}.  This facilitates theoretical predictions and their comparisons~\cite{2014qinqin,2010chy} to experimental measurements.
However, the experimental measurements of $D\to\phi{P}$ are still limited~\cite{pdg} due to the relative low branching fractions (BF) which are suppressed by phase space due to the $\phi$ meson mass.  The singly Cabibbo-suppressed (SCS)  decays of $D^{+}\to \phi\pi^{+}$~\cite{2008phipip}, $D^{0}\to \phi\pi^{0}$~\cite{2007phipi0}, and $D^{0}\to\phi\eta$~\cite{2004phieta} have been studied by CLEO, BaBar and Belle, respectively. The BF of the doubly Cabibbo-suppressed (DCS) decay $D^{+}\to\phi  K^{+}$ is derived out according to two measurements of the total BF for $D^{+}\to{K^-K^+K^+}$ and the intermediate fraction of $\phi K^{+}$  at LHCb~\cite{2019phik,2019kkk}.

According to isospin symmetry between $u$ and $d$ quarks, the BFs for $D^{0}\to\phi\pi^{0}$ and $D^{+}\to\phi\pi^{+}$ are connected~\cite{2014qinqin,2010chy} as follows:
\begin{equation}
\frac{\BR(D^{0}\to\phi\pi^{0})}{\BR(D^{+}\to\phi\pi^{+})}=\frac{1}{2}\frac{\Gamma_{D^{+}}}{\Gamma_{D^{0}}}=\frac{1}{2}\frac{\tau_{D^{0}}}{\tau_{D^{+}}}.
\label{eq:isospin}
\end{equation}
However, the current experimental result for the BF ratio deviates from prediction value of Eq.~\eqref{eq:isospin} by $2.7\sigma$ as shown in Table~\ref{tab:isospin}. Therefore, improved measurement is necessary to further test it and help to understand the strong interaction in $D$ meson hadronic decays.

In this analysis, we study four two-body decay modes of $D\to\phi{P}$, which are $D^{+}\to\phi K^{+}$, $D^{+}\to\phi\pi^{+}$, $D^{0}\to\phi\pi^{0}$, and $D^{0}\to\phi\eta$, based on a data set of 2.93 fb$^{-1}$~\cite{luminosity} taken at $\sqrt{s} = 3.773$~\gev~with the BESIII detector.
Due to energy conservation, the $D$ and $\bar{D}$ mesons from $e^+e^- \to \psi(3770) \to D\bar{D}$ are always produced in a pair without any other accompanying hadrons.
Throughout this paper, charge-conjugate modes are implied.
\begin{tablehere}
\caption{Current result of the ratio of $\BR(D^0\to\phi\pi^0)$ to $\BR(D^+\to\phi\pi^+)$.}
\label{tab:isospin}
\centering
\begin{footnotesize}
\begin{tabular}{lcc}
\toprule
                    The ratio & Experiment result~(\%) & Prediction(\%) \\
                    \midrule
                    $\frac{\BR(D^{0}\to\phi\pi^{0})}{\BR(D^{+}\to\phi\pi^{+})}$ & $24.6\pm1.8$~\cite{2008phipip,2007phipi0} & $19.7\pm0.2$~\cite{lifetime}\\
\bottomrule
\end{tabular}
\end{footnotesize}
\end{tablehere}
\section{BESIII Detector and Monte Carlo Simulation}

The BESIII detector is a magnetic
spectrometer~\cite{Ablikim:2009aa} located at the Beijing Electron
Positron Collider (BEPCII)~\cite{Yu:IPAC2016-TUYA01}. The
cylindrical core of the BESIII detector consists of a helium-based
 multilayer drift chamber (MDC), a plastic scintillator time-of-flight
system (TOF), and a CsI(Tl) electromagnetic calorimeter (EMC),
which are all enclosed in a superconducting solenoidal magnet
providing a 1.0~T magnetic field. The solenoid is supported by an
octagonal flux-return yoke with resistive plate counter muon
identifier modules interleaved with steel. The acceptance of
charged particles and photons is 93\% of the $4\pi$ solid angle. The
charged-particle momentum resolution at $1~{\rm GeV}/c$ is
$0.5\%$, and the specific energy loss ($dE/dx$) resolution is $6\%$ for electrons
from Bhabha scattering. The EMC measures photon ~~energies with a
resolution of $2.5\%$ ($5\%$) at $1$~GeV in the barrel (end cap)
region. The time resolution of the TOF barrel section is 68~ps, while
that of the end cap is 110~ps.

Simulated samples produced with the {\sc
geant4}-based~\cite{geant4} Monte Carlo (MC) package, which
includes the geometric description~\cite{geo1,geo2} of the BESIII detector and the
detector response, are used to determine the detection efficiency
and to estimate the backgrounds. The simulation includes the beam
energy spread and initial state radiation (ISR) in the $e^+e^-$
annihilations modelled with the generator {\sc
kkmc}~\cite{ref:kkmc}. The inclusive MC samples consist of the production of $D\bar{D}$
pairs, the non-$D\bar{D}$ decays of the $\psi(3770)$, the ISR
production of the $J/\psi$ and $\psi(3686)$ states, and the
continuum processes incorporated in {\sc kkmc}~\cite{ref:kkmc}. The equivalent luminosity of the inclusive MC samples is about 10 times that of the data. The known decay modes are modelled with {\sc
evtgen}~\cite{ref:evtgen} using branching fractions taken from the
Particle Data Group~\cite{pdg}, and the remaining unknown decays
from the charmonium states with {\sc
lundcharm}~\cite{ref:lundcharm}. The final state radiations (FSR)
from charged final state particles are incorporated with the {\sc
photos} (version 2.02) package~\cite{photos,photos2}. The signal processes are generated separately taking the spin-matrix elements into account in {\sc evtgen}. For each signal channel, 200 000 events are simulated.

\section{Event Selection}

Candidates of the decay modes $D\to\phi P$ are reconstructed by combining the final states of $\kpm$, $\ppm$, $\piz$, and $\eta$ particles with BESIII offline software system~\cite{2009besphys,boss}, where $\phi$ mesons are detected via decays to $K^+K^-$. Candidates for $\piz$ and $\eta$ are identified from $\piz\to\gamma\gamma$ and $\eta\to\gamma\gamma$, respectively.

Selected charged tracks must satisfy $|\cos\theta| < 0.93$, where $\theta$ is the polar angle with respect to the beam axis. The distance of closest approach of the track to the interaction point is required to be less than 10 cm in the beam direction and less than 1 cm in the plane perpendicular to the beam. Separation of charged kaons from charged pions is implemented by combining the energy loss ($\dedx$) in the MDC and the time-of-flight information from the TOF. We calculate the probabilities $P(K)$ and $P(\pi)$ with the hypothesis of $K$ or $\pi$, and require that $K$ candidates have $P(K)>P(\pi)$, while $\pi$ candidates have $P(\pi)>P(K)$.

Photon candidates are selected from neutral showers deposited in the EMC crystals, with energies larger than 25 MeV in the barrel ($|\cos{\theta}|<0.8$) and 50 MeV in the end cap ($0.86<|\cos{\theta}|<0.92$). To reduce fake photons due to beam background or electronic noise, the shower clusters are required to be within [0, 700] ns from the event start time. Furthermore, the photon candidates are required to be at least $10^{\circ}$ away from any charged tracks to remove fake photons caused by the interactions of hadrons in the EMC.

The~$\pi^{0}\,(\eta)$~candidates are formed with pairs of photon candidates, whose invariant mass,~$M_{\gamma\gamma}$,~is required to be within [0.115, 0.150] ([0.500, 0.560])~\gev/$c^{2}$. To improve momentum resolution, a 1C kinematic fit constraining the reconstructed $\pi^0(\eta)$ mass to the nominal mass~\cite{pdg} is performed and the fitted four-momentum of the $\pi^{0}(\eta)$ is used in further analysis.

\section{Data Analysis}

In the rest frame of the initial $e^+e^-$ system, the total collision energy is shared equally by the $D\bar{D}$ pair. Hence, in this frame two variables, the energy difference $\Delta{E}$ and the beam constrained mass $M_{\rm BC}$ related to energy and momentum conservation, respectively, are defined as
\begin{align*}
& \Delta{E}\equiv E_{D}-\sqrt{s}/2, \\
&M_{\rm BC}\equiv \sqrt{s/4 c^{4}-|\vec{p}_{\rm D}|^{2}/c^{2}},
\end{align*}
where $\vec{p}_{\rm D}$ is the momentum of the $D$ candidate.

Signals for the four $\phi P$ decay modes are expected to peak around zero in $\Delta{E}$ distributions and the $D$ nominal mass in $M_{\rm BC}$ distributions. To suppress combinatorial background, the $\Delta{E}$ of the $D$ candidates are required to be within the regions listed in Table~\ref{tab:fitresult} for the different signal modes, which correspond to about $3\sigma$ coverage. The asymmetric boundaries of the $\Delta{E}$ region for the $\phi\pi^0$ and $\phi\eta$ modes are due to energy leakage in the EMC when reconstructing the photon energy. If there is more than one $D$ candidate left for one signal decay mode in an event, the candidate with the smallest $|\Delta{E}|$ is chosen for further analysis. More than $60\%$ of events have multiple candidates for $D^0$ decay modes and $20\%$ for $D^+$ decay modes.  According to the studies on MC samples, the probability to select the correct candidate by choosing the minimum $|\Delta{E}|$ is more than $90\%$. In addition, the credibility of this method is verified and proven to be robust by studying the high-statistics inclusive MC samples.

\begin{table*}[tp]
\centering
\caption{For each signal mode, the requirement on $\Delta{E}$, signal yields $N_{\rm sig}^i$, MC-determined detection efficiency $\varepsilon^i$, branching fraction $\BR^i$ in this work, and the corresponding world results $\BR_{\rm ext}$. }
\label{tab:fitresult}
\begin{footnotesize}
\begin{tabular}{lccccc}
\toprule
Decay mode              &$\Delta{E}$(\gev)  & $N_{\rm sig}^i$ & $\varepsilon^i$(\%) & $\BR^i(\times10^{-4})$  & $\BR_{\rm ext}(\times10^{-4})$
\\
\hline
$D^{+}\to\phi\pi^{+}$   &$[-0.020,0.019]$     & $17527\pm152$  & $37.7\pm0.1$        & $57.0\pm0.5\pm1.3$    & $53.7\pm2.3$~\cite{pdg}\\
\multirow{2}{*}{$D^{+}\to\phi{K^{+}}$} &
\multirow{2}{*}{$[-0.019,0.018]$ } &
\multirow{2}{*}{$12^{+28}_{-12}$ } &
\multirow{2}{*}{$23.7\pm0.1$ } &
$0.062^{+0.144}_{-0.062}\pm0.002$ &
\multirow{2}{*}{$0.085\pm0.011$~\cite{pdg,2019phik,2019kkk}} \\
~ & ~ & ~ & ~  & $<0.21$ at $90\%$ CL & ~\\
$D^{0}\to\phi\pi^{0}$   &$[-0.077,0.035]$     & $3333\pm76$    & $27.7\pm0.1$        & $11.68\pm0.28\pm0.28$  & $13.2\pm0.8$~\cite{pdg} \\
$D^{0}\to\phi\eta$      &$[-0.040,0.038]$     & $102\pm26$     & $13.7\pm0.1$        & $1.81\pm0.46\pm0.06$  & $1.4\pm0.5$~\cite{pdg}\\
\bottomrule
\end{tabular}
\end{footnotesize}
\end{table*}
\begin{figure*}[tp]
\centering
\subfigure[$D^+\to\phi\pi^+$]{
\includegraphics[width=3.7cm]{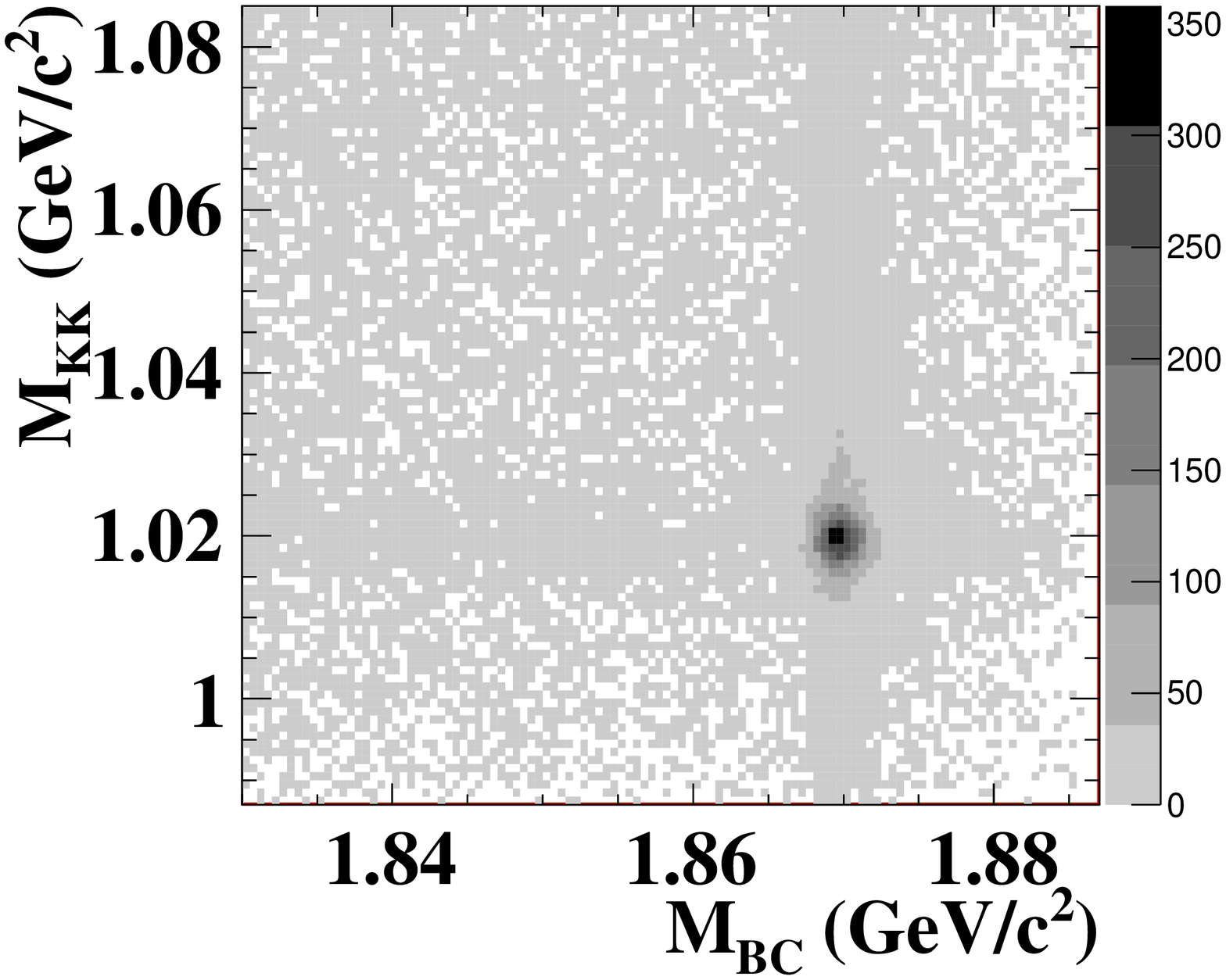}
}
\subfigure[$D^+\to\phi{K^+}$]{
\includegraphics[width=3.7cm]{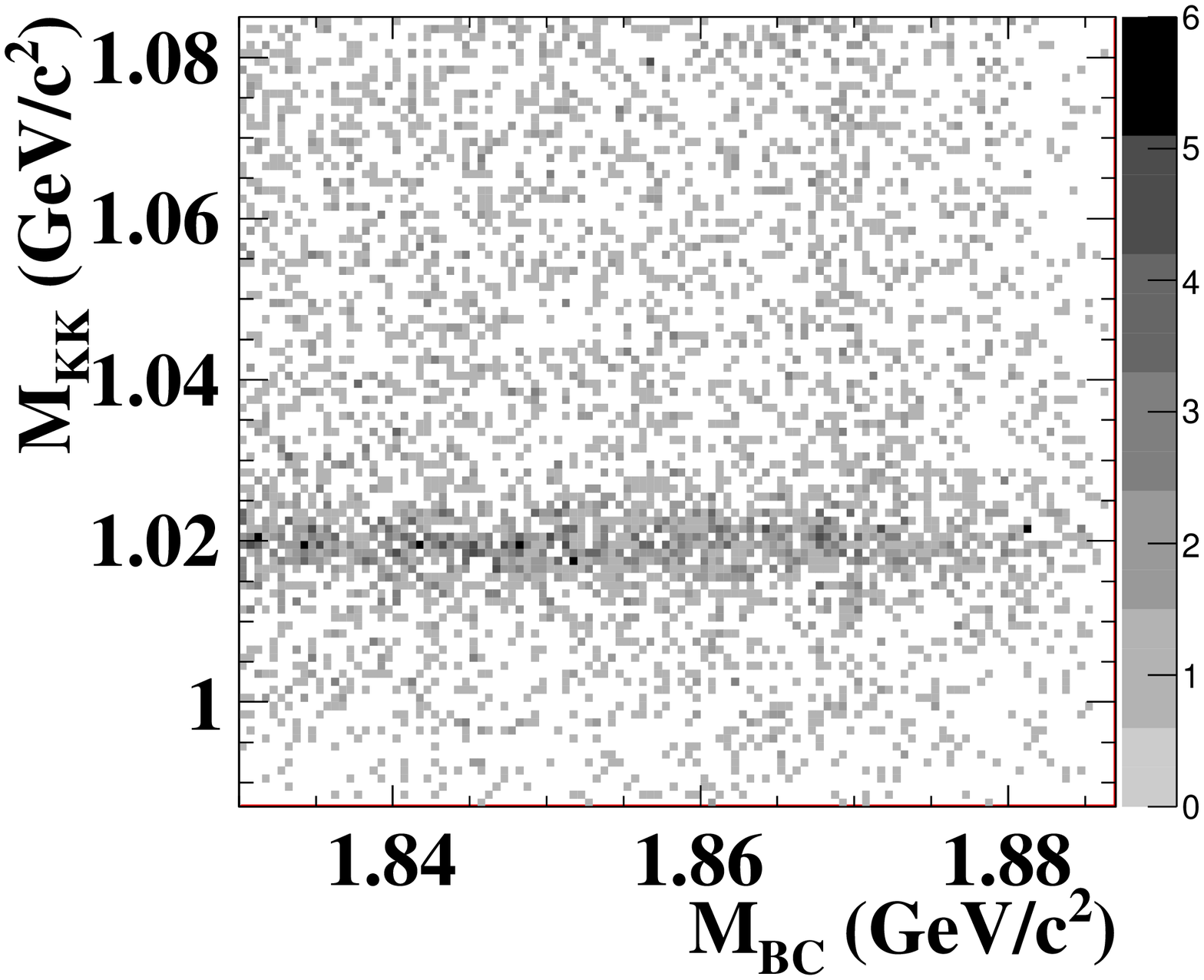}
}
\quad
\subfigure[$D^0\to\phi\pi^0$]{
\includegraphics[width=3.7cm]{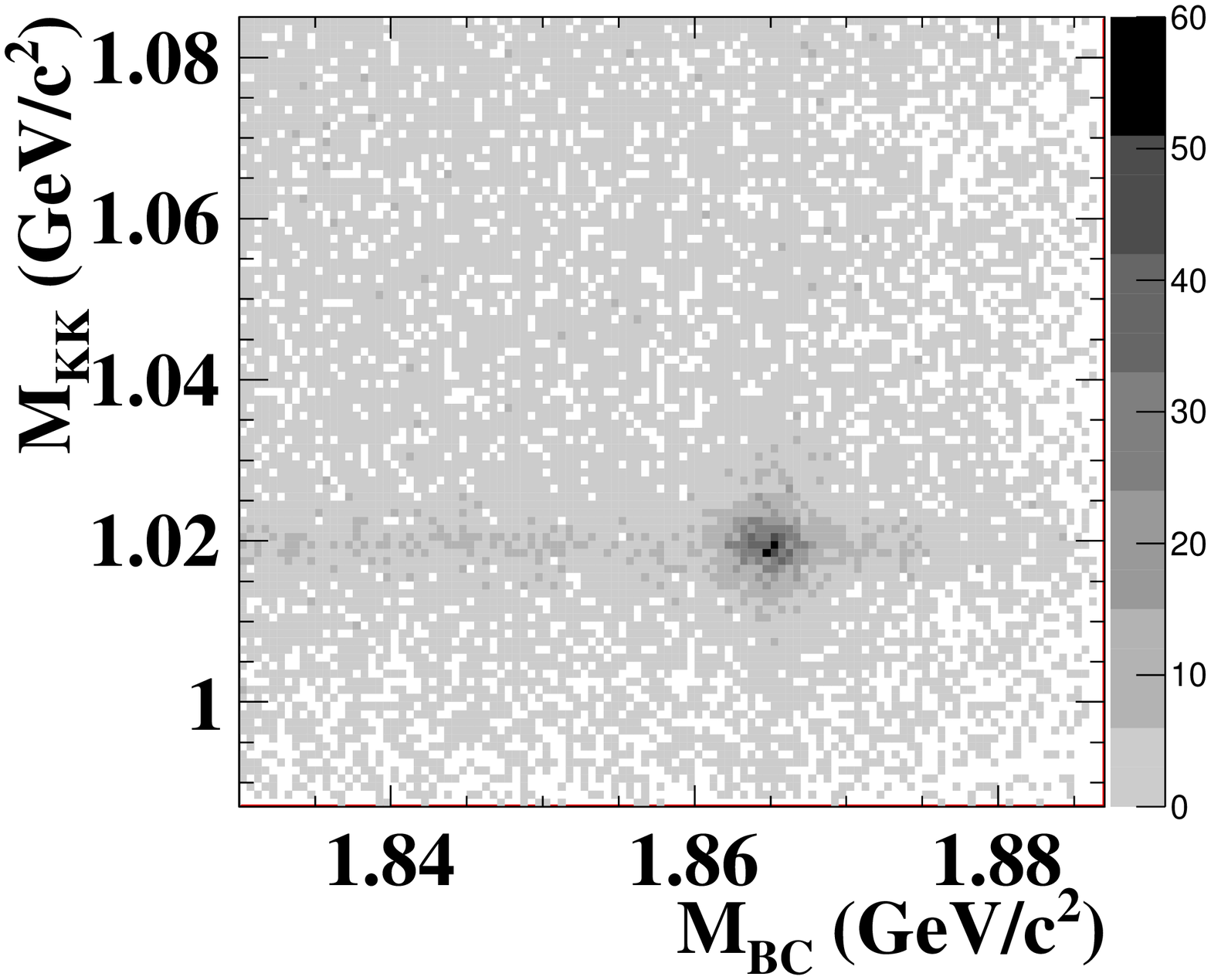}
}
\subfigure[$D^0\to\phi\eta$]{
\includegraphics[width=3.7cm]{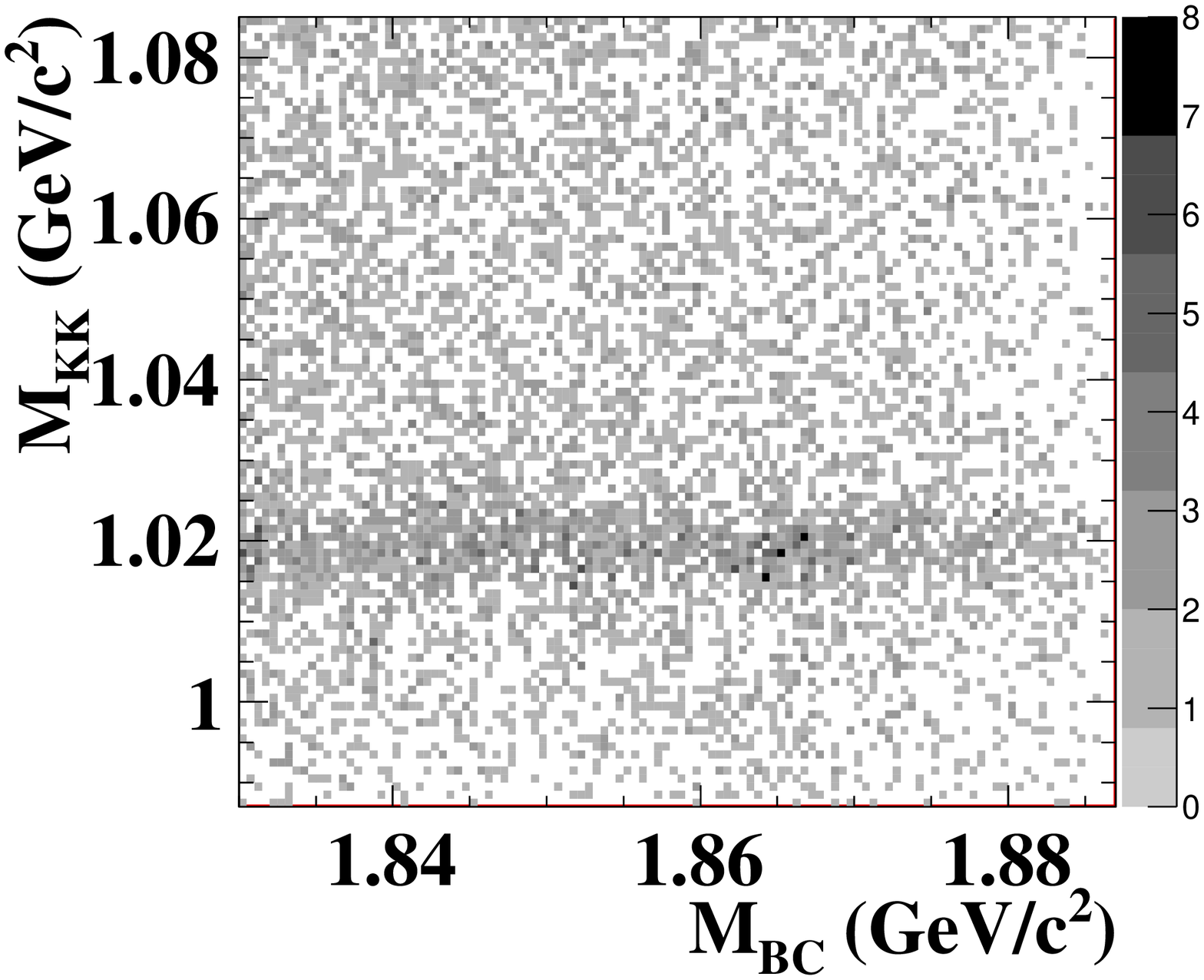}
}
\caption{
 Two-dimensional distributions of $M_{\rm BC}$ and $M_{\rm KK}$ in data for the four signal modes.
}
\label{fig:colz}
\end{figure*}
\begin{figure*}[tp]
\centering
\subfigure[$D^+\to\phi\pi^+$]{
\includegraphics[width=7.5cm]{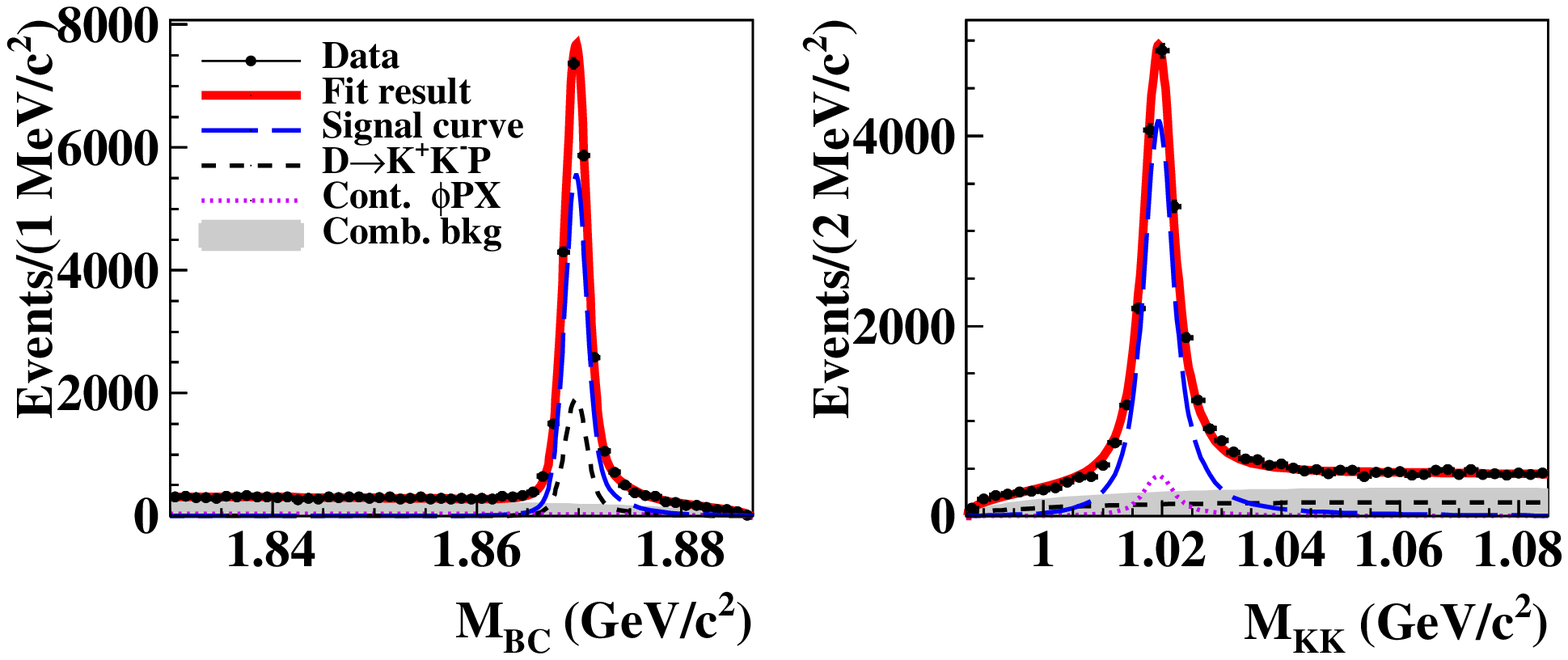}
}
\quad
\subfigure[$D^+\to\phi{K^+}$]{
\includegraphics[width=7.5cm]{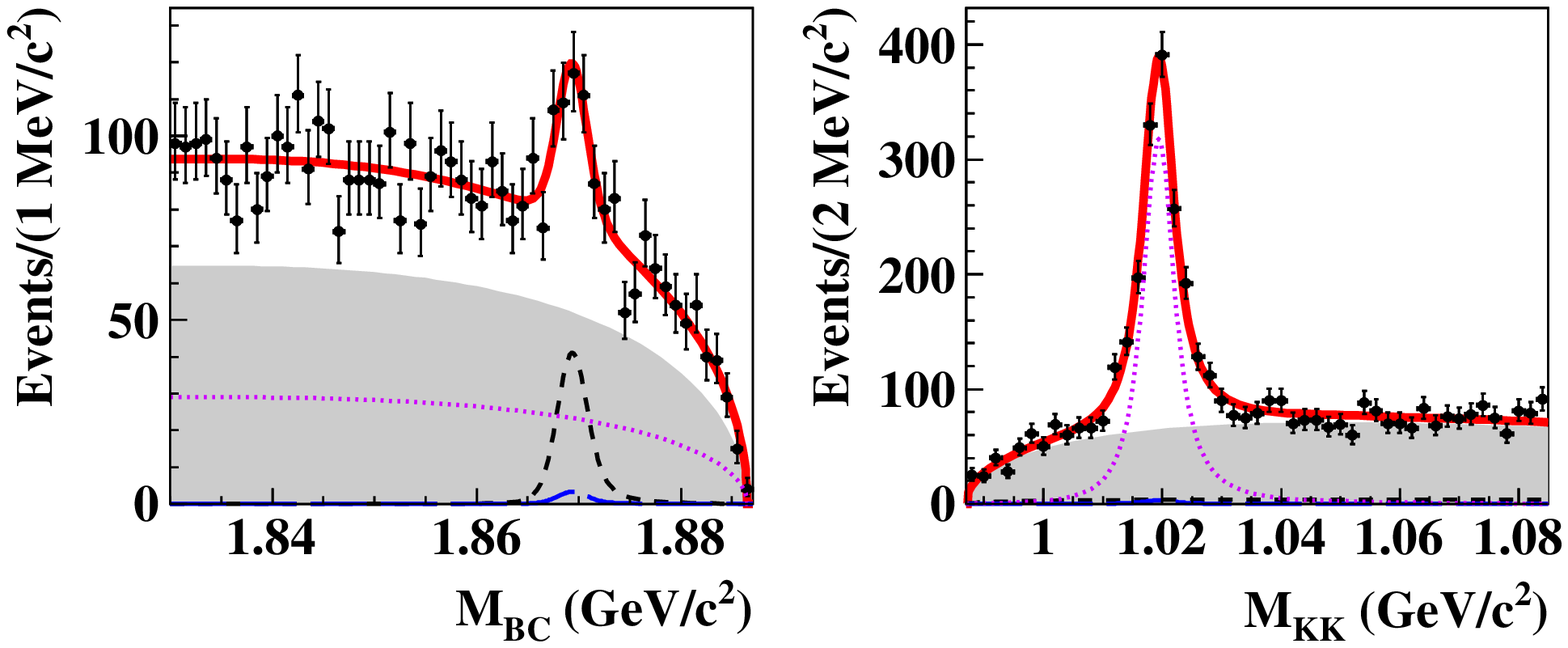}
}
\quad
\subfigure[$D^0\to\phi\pi^0$]{
\includegraphics[width=7.5cm]{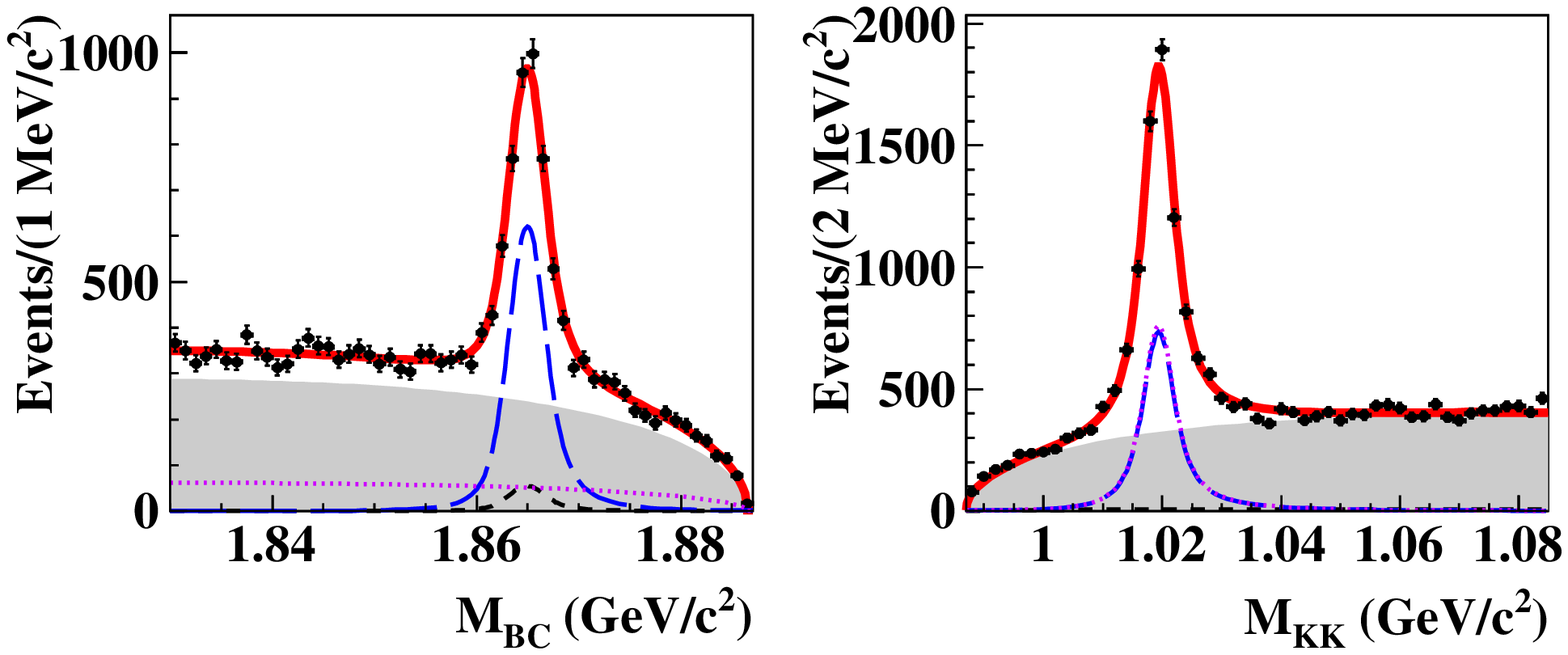}
}
\quad
\subfigure[$D^0\to\phi\eta$]{
\includegraphics[width=7.5cm]{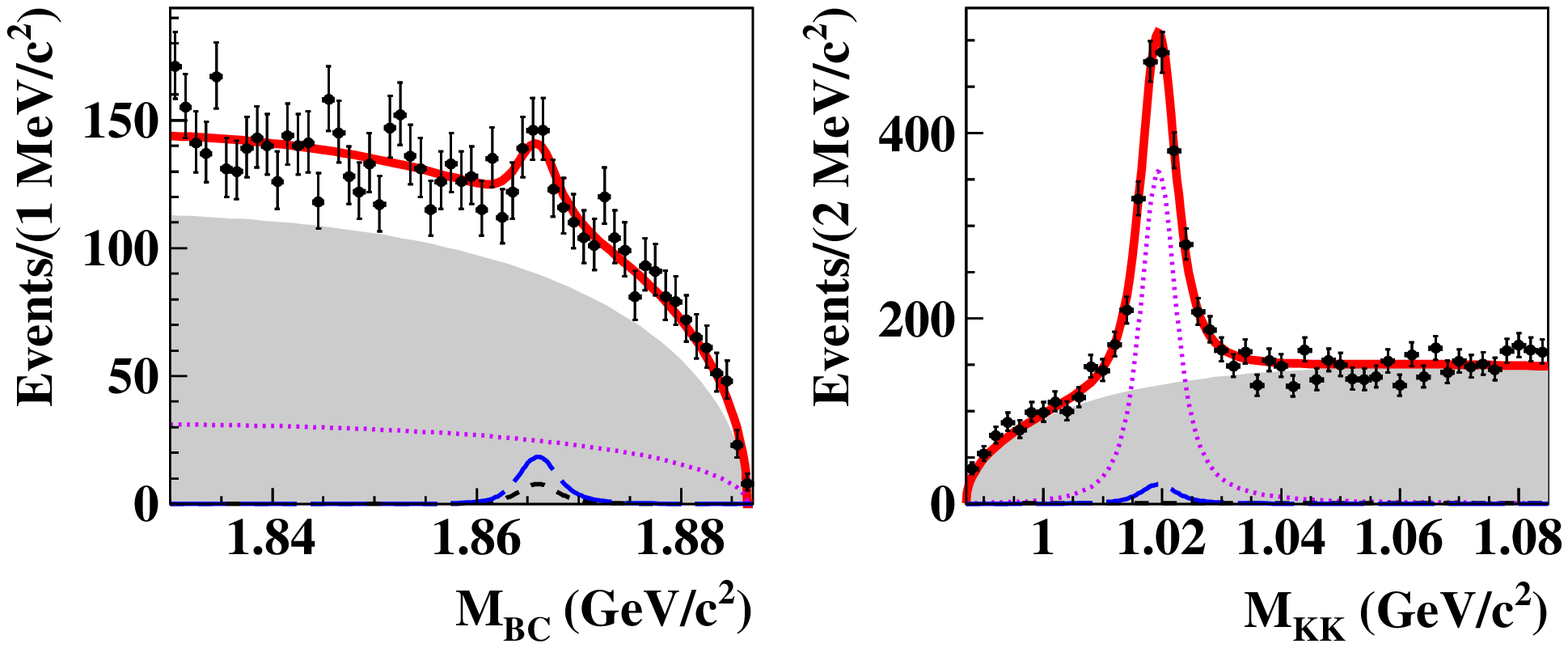}
}
\caption{
(Color online) Two-dimensional unbinned maximum likelihood fits to the distributions of $M_{\rm BC}$ and $M_{\rm KK}$ in data for the four signal modes.
The points with error bars are data,
the (red) thick curves are the total fits,
the (blue) long dashed curves describe the signals,
the (violet) dotted curves represent backgrounds of true $\phi$ mesons not from $D\to\phi P$ decay modes, the (black) dashed curves describe  backgrounds from $D\to K^+K^-P$ without a $\phi$ meson, and the shaded area show the combinatorial backgrounds.
}
\label{fig:fitresult}
\end{figure*}

As shown in Fig.~\ref{fig:colz} and Fig.~\ref{fig:fitresult}, clear peaks are seen in the $M_{\rm BC}$ and $M_{\rm KK}$ distributions for the four signal modes, which correspond to the $D\to K^+K^- P$ signals and $\phi\to K^+K^-$ signals, respectively. According to the studies based on the inclusive MC samples, three types of background events will pass through above selection criteria. The first one is a true $D$ meson decaying to $K^+K^-P$ final states without a $\phi$ meson involved ($D\to{K^+K^-}P$), the second one is a true $\phi$ meson not from the corresponding signal mode (Cont. $\phi{PX}$) and the third one is the combinatorial background from neither of the previous two sources (Comb. bkg).

Two-dimensional unbinned extended maximum likelihood fits to the obtained distributions of $M_{\rm BC}$ and $M_{\rm KK}$ are performed to extract yields of signals, as shown in Fig.~\ref{fig:fitresult}. The $M_{\rm KK}$ variable is employed here to discriminate the $\phi$ meson signal from the non-resonant $K^+K^-$ final state. The probability density functions of the $D$ meson and $\phi$ meson signals are modeled by the MC-simulated signal shapes convoluted with Gaussian functions that describe the resolution differences between MC simulations and data. The combinatorial backgrounds in $M_{\rm BC}$ ($M_{\rm KK}$) are described with (inverted) ARGUS~\cite{Albrecht:1990am} functions based on the studies on the inclusive MC sample. Since the correlation between $M_{\rm KK}$ and $M_{\rm BC}$ can be neglected, these two variables are considered uncorrelated in the fit. The parameters of the (inverted) ARGUS and Gaussian functions in two-dimensional fits are fixed according to one-dimensional fits to the corresponding $M_{\rm BC}$ and $M_{\rm KK}$ distributions. The obtained signal yields are given in Table~\ref{tab:fitresult}.

\section{Branching Fraction}

The branching fractions for the $D\to\phi P$ decays can be calculated by
\begin{eqnarray}
\BR^i=\frac{N_{\rm sig}^i}{2\cdot{N_{D\bar{D}}}\cdot\varepsilon^i\cdot{\cal B}^i_{\rm sub}},
\label{eq:bf}
\end{eqnarray}
where $i$ denotes a signal mode of $D\to\phi P$, $N_{\rm sig}^i$ is the signal yield extracted in data, $N_{D\bar{D}}$ is the number of $D\bar{D}$ event in data, which is $(8296\pm31\pm64)\times10^{3}$ for $D^{+}D^{-}$ and $(10597\pm28\pm89)\times10^{3}$ for $D^{0}\bar{D}^{0}$~\cite{2014ndd} in the data set we analyzed, $\varepsilon^i$ is the reconstruction efficiency determined from MC simulation of the signal mode, and $\BR^i_{\rm sub}$ are the branching fractions of the intermediate decay processes $\phi\to{K^+K^-}$ and $\piz/\eta\to\gamma\gamma$, quoted from PDG~\cite{pdg}. The branching fraction for each decay mode is calculated in Table~\ref{tab:fitresult}.

\begin{figurehere}
  \centering
  \includegraphics[width=6cm]{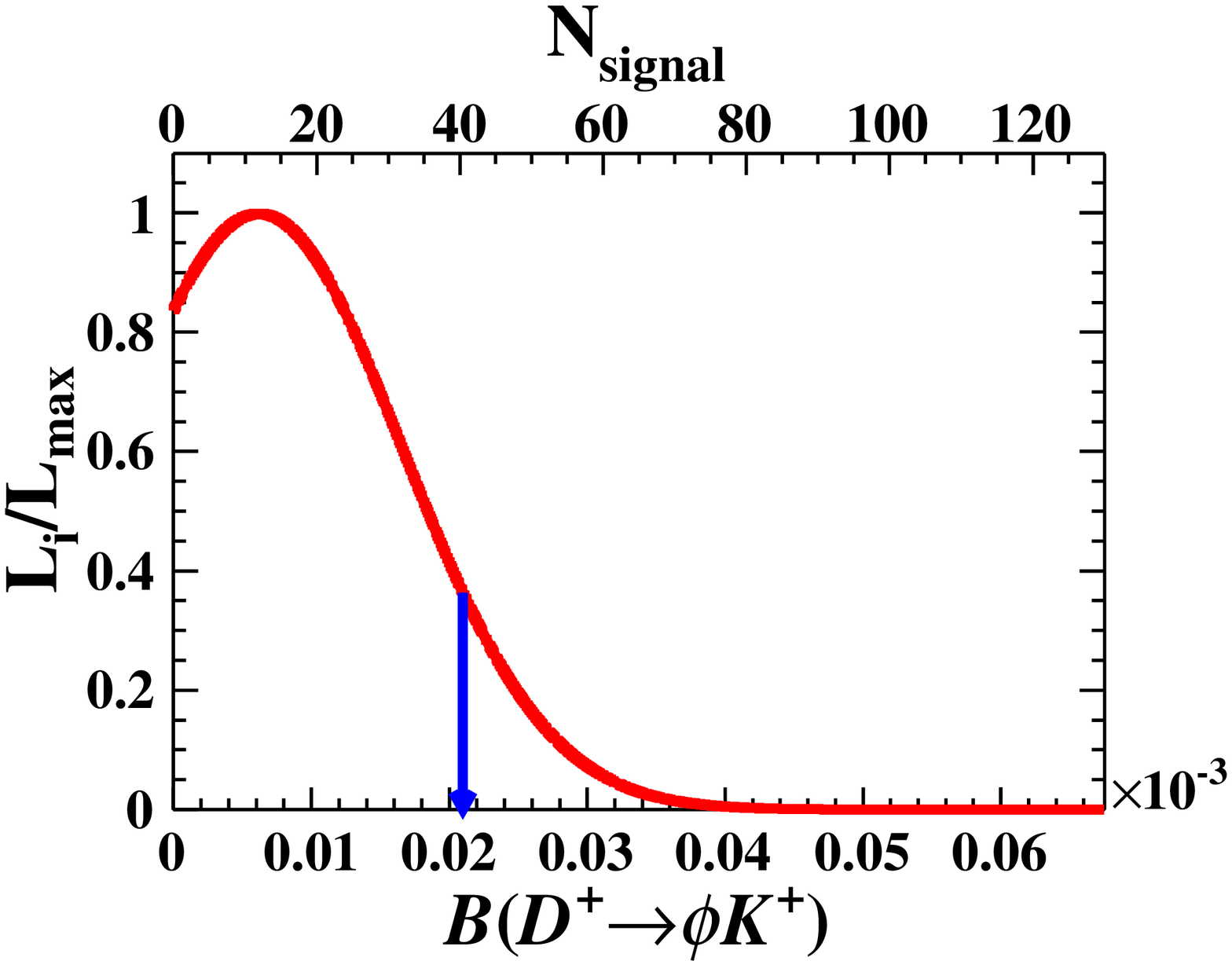}
  \caption{Likelihood curve as the function of assumed $\BR(D^{+}\to\phi{K^{+}})$. The arrow points to the position of upper limit at the $90\%$ CL.}
  \label{fig:smear}
\end{figurehere}

The statistical significance of the $D^0\to\phi\eta$ signal is evaluated, taken as $\sqrt{-2\ln(\mathcal{L}_{0}^{\rm stat}/\mathcal{L}_{\rm max}^{\rm stat})}$ where $\mathcal{L}_{\rm max}^{\rm stat}$ and $\mathcal{L}_0^{\rm stat}$ are the maximum likelihood values with and without signal, respectively, to be $4.2\sigma$. Since the significance of the observed $D^+\to\phi K^+$ signal is $0.8\sigma$, the upper limit of $\BR(D^+\to\phi K^+)$ is estimated by a likelihood scan method, which takes into account the systematic uncertainties as follows
\begin{eqnarray}
\mathcal{L}_i(\BR^i)=\int^{1}_{-1}\mathcal{L}^{\rm stat}[(1+\Delta)\BR^i]\exp\left(-\frac{\Delta^2}{2\sigma^{2}_{i,{\rm syst}}}\right)\,d\Delta.
\label{eq:likelihood}
\end{eqnarray}
Here, $\Delta$ is the relative deviation of the estimated branching fraction from the nominal value and $\sigma_{i,{\rm syst}}$ is the total systematic uncertainty given in Table~\ref{tab:sys_err}.

The likelihood curve calculated according to Eq.~\eqref{eq:likelihood} is shown in Fig.~\ref{fig:smear}. The upper limit on $\BR(D^+\to\phi{K^+})$ at the $90\%$ confidence level (CL) is estimated to be  $2.1\times10^{-5}$ by integrating the likelihood curve in the physical region, $\BR^i>0$.

\begin{table*}[tp]
\caption{Summary of systematic uncertainties in percentage.}
\centering
\footnotesize
\begin{tabular}{lcccccccccc}
\toprule
			Source & $D^+\to\phi\pip$ & $D^+\to\phi K^{+}$ & $D^{0}\to\phi\pi^{0}$ & $D^{0}\to\phi\eta$ & $\frac{\BR(D^{0}\to\phi\pi^{0})}{\BR(D^{+}\to\phi\pi^{+})}$ \\
			\hline
			Tracking                               &$1.0$  &$1.1$  &$0.8$  &$1.0$  & $0.3$\\
			PID                                    &$1.2$  &$1.0$  &$0.6$  &$0.6$  & $0.4$\\
			$\pi^{0}$ reconstruction               &$-$    &$-$    &$1.2$  &$-$    & $1.2$\\
			$\eta$ reconstruction                  &$-$    &$-$    &$-$    &$1.8$  & $-$\\
			$\Delta E$ requirement                 &$0.2$  &$0.2$  &$0.2$  &$0.2$  & $0.3$\\
			2D fit                                 &$0.4$  &$2.5$  &$0.4$  &$2.0$  & $0.6$\\
			$N_{D\overline{D}}$ uncertainty        &$0.9$  &$0.9$  &$0.9$  &$0.9$  & $1.3$\\
			$\BR(\phi \to K^{+} K^{-})$            &$1.0$  &$1.0$  &$1.0$  &$1.0$  & $-$\\
			$\BR(\pi^{0}, \eta \to \gamma\gamma)$  &$-$    &$-$    &$0.1$  &$0.5$  & $0.1$\\
			QC effect                              &$-$    &$-$    &$1.0$  &$1.0$  & $1.0$\\
			\hline
                         Total                     &$2.2$  &$3.3$  &$2.4$  &$3.5$  & $2.2$ \\
\bottomrule
\end{tabular}
\label{tab:sys_err}
\end{table*}

\section{Systematic Uncertainties}\label{section:sys}

The following sources of systematic uncertainties, as given in Table~\ref{tab:sys_err}, are considered. The total systematic uncertainty is determined by adding all contributions in quadrature.

The uncertainties of tracking and particle identification (PID) for charged kaon and pion mesons, as well as $\pi^0(\eta)$ reconstruction, have been studied in previous works by using control samples of $D$ hadronic events~\cite{2017trackpid}. The uncertainties are weighted according to the kinematics of the candidates.  Furthermore, in order to estimate the systematic uncertainty caused by the selected $\pi^0(\eta)$ signal regions, the requirements on $M_{\gamma\gamma}$ are varied and the resultant changes on the BFs are $0.7\%$ ($1.1\%$). This uncertainty is combined with that of  $\pi^0(\eta)$ reconstruction, the quadrature sum of which is given as $1.2\% (1.8\%)$. Requirements on $\Delta E$ are studied by smearing the corresponding $\Delta E$ distribution in inclusive MC samples with Gaussian functions and re-calculating  detection efficiencies. The changes of the efficiencies are assigned as the corresponding uncertainties.

Systematic uncertainty related to the two-dimensional fit includes parameters of Gaussian and ARGUS functions, fit range and background models. For the fixed parameters in the Gaussian and ARGUS functions, their values are varied by $\pm1\sigma$ from the one-dimensional fit results and the largest resultant change is assigned as the systematic uncertainty. The uncertainty due to the fit range is estimated by repeating the fits with a series of varied ranges and the corresponding changes are found to be negligible. For the background models,  potential background of $D\to f_0(980) P$ is included in the fit and the change on the number of signal events is assigned as uncertainty. This uncertainty is larger for $\BR(D^+\to\phi{K^+})$ and $\BR(D^0\to\phi\eta)$ due to the smaller signal yields.

The uncertainties of the quoted $N_{D\bar{D}}$ from Ref.~\cite{2014ndd}, $\BR(\phi\to K^{+}K^{-})$ and $\BR(\pi^0/\eta \to \gamma \gamma)$
from PDG~\cite{pdg} are taken into account for the relevant signal modes.
Since $D^0$ and $\overline D^0$ are coherently produced in the process $e^{+}e^{-}\to \psi(3770)\to D^0\overline D^0$, quantum coherence (QC)~\cite{2006qc1} should be considered according to the equation
\begin{displaymath}
\Delta N^{obs}_{CP} = y_{CP}\cdot N^{obs}_{CP}.
\end{displaymath}
The uncertainty depends on the $D^0-{\overline D^0}$ mixing parameter $y_{CP}$,
and is taken to be $1.0\%$~\cite{y} conservatively.

For the systematic uncertainties of $\frac{\BR(D^{0}\to\phi\pi^{0})}{\BR(D^{+}\to\phi\pi^{+})}$, the effects related to $K^\pm$ tracking and PID are mostly cancelled, owing to their same kinematic phase space. The remaining systematic uncertainties in Table~\ref{tab:sys_err} are considered independently and summed up in quadrature.

\section{Summary}

The decays of $D^{+}\to\phi\pi^{+}$, $D^{0}\to\phi\pi^{0}$, $D^{0}\to\phi\eta$, and $D^{+}\to\phi K^{+}$ are studied by analyzing $2.93~\rm fb^{-1}$ data taken at $\sqrt{s} = 3.773$~\gev~with the BESIII detector. The obtained BFs are consistent with previous results, as listed in  Table~\ref{tab:fitresult}, while the precisions of the BFs for the first three modes are improved.
In addition, the upper limit on $\BR(D^{+}\to\phi K^{+})$ of $2.1\times10^{-5}$ at $90\%$ CL is reported.

Our results of $\BR(D\to\phi\pi)$ and $\BR(D^0\to\phi\eta)$ are consistent with the previous measurements. Furthermore, the ratio of $\BR(D^0\to\phi\pi^0)$ to $\BR(D^+\to\phi\pi^+)$ is calculated to be $(20.49\pm0.50\pm0.45)\%$, which is smaller than the previous result $(24.6\pm1.8)\%$~\cite{2008phipip,2007phipi0}. Meanwhile, the deviation from the predicted value of $(19.7\pm0.2)\%$ in Eq.~\eqref{eq:isospin} is reduced from $2.7\sigma$ to $1.2\sigma$,
which shows better agreement than the previous measurement.
Hence, our results support the isospin symmetry between these two $D$ meson decay modes.

\section{Acknowledgments}

The BESIII collaboration thanks the staff of BEPCII and the IHEP computing center for their strong support. This work is supported in part by National Key Basic Research Program of China under Contract No. 2015CB856700; National Natural Science Foundation of China (NSFC) under Contracts Nos. 11625523, 11635010, 11735014, 11822506, 11835012; the Chinese Academy of Sciences (CAS) Large-Scale Scientific Facility Program; Joint Large-Scale Scientific Facility Funds of the NSFC and CAS under Contracts Nos. U1532257, U1532258, U1732263, U1832207; CAS Key Research Program of Frontier Sciences under Contracts Nos. QYZDJ-SSW-SLH003, QYZDJ-SSW-SLH040; 100 Talents Program of CAS; INPAC and Shanghai Key Laboratory for Particle Physics and Cosmology; ERC under Contract No. 758462; German Research Foundation DFG under Contracts Nos. Collaborative Research Center CRC 1044, FOR 2359; Istituto Nazionale di Fisica Nucleare, Italy; Koninklijke Nederlandse Akademie van Wetenschappen (KNAW) under Contract No. 530-4CDP03; Ministry of Development of Turkey under Contract No. DPT2006K-120470; National Science and Technology fund; STFC (United Kingdom); The Knut and Alice Wallenberg Foundation (Sweden) under Contract No. 2016.0157; The Royal Society, UK under Contracts Nos. DH140054, DH160214; The Swedish Research Council; U. S. Department of Energy under Contracts Nos. DE-FG02-05ER41374, DE-SC-0010118, DE-SC-0012069; University of Groningen (RuG) and the Helmholtzzentrum fuer Schwerionenforschung GmbH (GSI), Darmstadt

\end{multicols}
\end{document}